\documentclass[a4paper,conference]{IEEEtran}
\usepackage{times}  
\usepackage{helvet}  
\usepackage{courier}  
\usepackage{url}  
\usepackage{graphicx}  
\usepackage{tabularx}
\usepackage{amssymb}
\usepackage{amsfonts}
\usepackage{mathrsfs}
\usepackage{booktabs}
\usepackage{xcolor}
\usepackage{multirow}
\usepackage{xargs}
\usepackage{bm}
\usepackage{graphicx}
\usepackage{stfloats}
\usepackage{pgfplots}
\usepackage{nicefrac}

\usepackage{amsmath}

\graphicspath{{./figs/}}

\usepackage{cleveref}

\usepackage{tikz}

\usepackage{subcaption}
\usepackage{amsthm}

\usepackage{etoolbox}

\newtheorem{theorem}{Theorem}

\begin{document}


\date{}

\title{Killing four birds with one Gaussian process: \\ 
the relation between different test-time attacks}

\author{\IEEEauthorblockN{Kathrin Grosse}
\IEEEauthorblockA{CISPA \\
Saarland Informatics Campus}
\and
\IEEEauthorblockN{Michael T. Smith}
\IEEEauthorblockA{Department of Computer Science \\
University of Sheffield}
\and
\IEEEauthorblockN{Michael Backes}
\IEEEauthorblockA{CISPA Helmholtz Center \\
for Information Security}}

\maketitle

\begin{abstract}%
In machine learning (ML) security, attacks like evasion, model stealing or membership inference are generally studied in individually.
Previous work has also shown a relationship between some attacks and decision function curvature of the targeted model. 
%
%
%
Consequently, we study an ML model allowing direct control over the decision surface curvature: Gaussian Process classifiers (GPCs).
For evasion, we find that changing GPC's curvature to be robust against one attack algorithm boils down to enabling a different norm or attack algorithm to succeed. 
This is backed up by our formal analysis showing that static security guarantees are opposed to learning.
Concerning intellectual property, we show formally that lazy learning does not necessarily leak all information when applied.
In practice, often a seemingly secure curvature can be found.
For example, we are able to secure GPC against empirical membership inference by proper configuration. 
In this configuration, however, the GPC's hyper-parameters are leaked, e.g. model reverse engineering succeeds. 
We conclude that attacks on classification should not be studied in isolation, but in relation to each other.
\end{abstract}
%
%
%
%

%
%
%

\section{Introduction}

Security researchers study a plethora of attacks on machine learning (ML).
In general, each attack is studied individually. For example, 
\textit{Evasion attacks}, or \textit{adversarial examples}, are small perturbations added to a sample, which is subsequently misclassified. Examples for targeted systems include, but are not limited, to Malware detectors~\cite{DBLP:conf/sp/SrndicL14,DBLP:journals/corr/DemontisMBMARCG17}, vision for autonomous driving that misclassifies traffic signs~\cite{2018arXiv180206430S}, 
and robot visual systems~\cite{DBLP:conf/iccvw/MelisDB0FR17}. Securing models against evasion was shown to lead to an arms race~\cite{DBLP:journals/corr/CarliniW17,athalye2018obfuscated}. 

Some attacks harm the intellectual property of the model owner. In \textit{model stealing}~\cite{DBLP:journals/corr/PapernotMG16,DBLP:conf/uss/TramerZJRR16}, the attacker copies a model's functionality without consent of the owner. In \textit{model reverse engineering}~\cite{joon18icrl}, hyper parameters of the model are inferred illegitimately. 
Another attack retrieves the data that was used to train the classifier, called \textit{membership inference} \cite{2017arXiv170207464H,DBLP:journals/corr/abs-1806-01246}. This corresponds to a privacy breach for the subjects in the data, and/or a financial loss for the owner of the data.

A relationship between decision function curvature and membership inference was shown in~\cite{2017arXiv170207464H}.
An analogous relationship for
evasion has been found in linear models like support vector machines~\cite{DBLP:conf/ccs/RussuDBFR16} and used for mitigations in deep neural networks~\cite{2017arXiv170508475H,raghunathan2018certified}. 
These findings raise several questions. 
Are all test-time attacks related to decision function curvature? 
Do changes in curvature have the same effect on all attacks? 
To answer these questions, we need a model where curvature can be configured.
Such a model are Gaussian processes (GPs): 
Choosing a long \textit{lengthscale} before training yields for example a GP with a flat decision surface.

Studying GP yields two more benefits. GP are often applied in medical settings~\cite{CHEN200759,doi:10.1177/0272989X03261561,6238311}. Risk assessment for leaked data or learned parameters is thus crucial.  
Furthermore, GPs provide the means for a rigorous analysis: After training, a GP yields a closed form expression, where classification depends directly on both parameters learned and used training data.

\textbf{Contributions.} 
Our formal analysis confirms vulnerability towards evasion at test time once the GP has learned. Due to its mathematical form, GP allows to analytically compute the lengthscale iff the training data is known and only one lengthscale used.
We further conduct a broad empirical study of vulnerability on six data sets, focusing on decision function curvature. 
To this end, we introduce two model reverse engineering attacks, one for GP's lengthscale, one for the kernel.
Decision function curvature often only changes the kind of attack that succeeds. In evasion, highly optimized attacks tend to fool a steep curvature. This steep curvature also leaks the data.
On the other hand, one-step evasion attacks are more effective on flat curvature.
This flat curvature also leaks parameters like the lengthscale.
In contrast, leakage of the kernel occurs at any lengthscale.
We conclude that attacks on classification should not be studied in isolation: mitigating one attack might just enable a different attack.


\section{Related Work}
To the best of our knowledge, few works have studied the relationship between  different attacks. 
Most works focus on deep learning, and on at most two attacks.
For example Suciu et al.~\cite{suciu2018does} study evasion and training time attacks jointly. Song and Mittal~\cite{MIAdvTRain} show that neural networks that are robust against evasion are more vulnerable against membership inference. Along these lines, there are defenses taking into account several attacks on deep learning~\cite{juuti2018prada,chou2018sentinet}. 
We instead focus on an in depth study of the \emph{relationship} between several attacks, and are unaware of any similar work.

Most formal works of test-time attacks focus on evasion~\cite{DBLP:journals/corr/WangJC17,2018arXiv180208686F,DBLP:journals/corr/TanayG16}. Our formal evasion analysis for GP is in the finite sample setting. Wang et al.~\cite{DBLP:journals/corr/WangJC17} instead give an analysis in the infinite sample limit on k-nearest-neighbors.  
A formal approach related to membership inference is the recent work on differential privacy for GP (see for example~\cite{DBLP:conf/aistats/SmithAZL18}). On the other hand, empirical evasion security has been studied on GP~\cite{2017arXiv170702476B,2017arXiv171106598G,bogunovic2018adversarially}. GPC also allows one to bound evasion vulnerability~\cite{blaas2019robustness,smith2019adversarial}.
We are not aware of any works studying model reverse engineering or model stealing on GPs.
\section{Background}
We introduce GP, give a short summary of adversarial learning and finally describe our threat model.

\subsection{Gaussian Process Classification}\label{sec:gpc}
We use Gaussian Process Classification (GPC)~\cite{DBLP:books/lib/RasmussenW06} for two classes using the Laplace approximation. The goal is to predict the labels $Y_t$ for the test data points $X_t$ accurately. 

We specify $k$ as covariance function or kernel and
 introduce GP regression (GPR). Assuming the data is produced by a GP,
\begin{equation}
\begin{bmatrix} Y_{\text{tr}} \\ Y_t \end{bmatrix} =  \mathcal{N} \left( 0,\begin{bmatrix} K_{\text{tr}} & K_{\text{tt}} \\K_{\text{tt}}^\top & K_t \end{bmatrix} \right) \text{   ,}
\label{Gaussian normal}
\end{equation}
where $Y_{\text{tr}}$ are the training labels, 
$K_{\text{tr}}$ is the covariance of the training data, $K_t$ of the test data, and $K_{\text{tt}}$ between test and training data. Having represented the data, we now review how to use this representation for predictions. As we use a Gaussian model, our predictions are Gaussian too, with a predictive mean and a predictive variance which we define now. At a given test point $x'$, assuming a Gaussian likelihood function, the predictive mean $y_{\text{t}}^*$ of the latent function on test data is 
\begin{equation}
y_{\text{t}}^* = K_{x'}^\top K_{\text{tr}}^{-1}Y_{\text{tr}} \text{   ,}
\label{pi}
\end{equation}
where $K_{x'}^\top$ is the vector with the covariances from test point $x'$ to each training point in $X_{tr}$. 

Classification, unlike regression, has binary outputs and requires a different likelihood and associated link function. We can approximate this
in a variety of ways. One of the simplest is the Laplace approximation whose simplicity allows us to formally analyze the GPC.
Finally, whenever we write GP, we refer to properties that both GPC and GPR share.

\subsection{Adversarial Machine Learning at Test Time}
We review all test time attacks dealt with in this paper: evasion, model reverse engineering, membership inference, and conclude with model stealing.

In \textbf{evasion}, the attacker computes a small perturbation $\delta$ for a trained classifier $f(X_t)=Y_t$  and a sample $x$ such that
\begin{equation}\label{def:advEx}
 \min \delta : f(x) \neq f(x+\delta) \text{    }
\end{equation}
the minimal $\delta$ changes the classification of $x$. There are two basic constraints here, one is to minimize $\delta$ to be as small as possible. We might, however, also maximize the confidence $f$ has in its classification, to be sure the resulting example is misclassified when targeting a second classifier $f'$. In general, we distinguish targeted and non-targeted attacks. Since we only evaluate on binary tasks, however, this distinction is superfluous in this paper.
Before we detail on the attacks used in this work, we address how to measure $\delta$. We might use the $L_0$ metric which counts the number of changed features. It is well suited for binary data, such as Malware features. The $L_2$ metric is equivalent to the euclidean or squared-root distance, and thus well suited for images. Another metric for images is the $L_\infty$ metric that measures the largest change introduced.
Many algorithms exist to craft adversarial examples. We now recap the algorithms used in our evaluation. The fast gradient sign method (\textbf{FGSM})~\cite{goodfellow2015explaining} is an untargeted one-step attack. We write  
\begin{equation*}
\delta =  \epsilon \times \text{sign}(\nabla_{x} J_L(F(x,\theta)))\text{    .}
\end{equation*}
In other words, a step adds the gradient of the model's loss w.r.t. the input $x'$ to the original sample.
The step size is parametrized by $\epsilon$.
FGSM minimizes the $L_\infty$ norm, as the same change is applied to all features. 
 The Jacobian-based saliency map approach (\textbf{JSMA})~\cite{papernot2016limitations} picks iteratively a pixel for perturbation that maximizes the output for the target class and minimizes the output for all other classes. This search is executed iteratively until miss-classification is achieved or the perturbation is too large. 

Finally, the $\mathbf{L_x}$ attacks~\cite{DBLP:journals/corr/CarliniW16a}  formulate evasion as an iterative optimization problem. 
The basis, or $L_2$ attack is formalized as the following optimization problem 
\begin{equation*}
 \underset{\delta}{\text{min}} \parallel 0.5(\tanh (\delta)+1)+x \parallel_2 + sg(0.5(\text{tanh}(\delta)+1)) \text{   ,}
\end{equation*} 
where $\tanh$ ensures that the box-constraint to enforce that no feature is set to higher values than in benign data. Term $s$ trades-off the constraint and function $g$. This function represents how confidently the network $f$ misclassifies $x+\delta$. Other variants of this attack minimize the $L_0$ or $L_\infty$ norms~\cite{DBLP:journals/corr/CarliniW16a}.


We now review attacks harming intellectual property (IP).

In \textbf{model reverse engineering}, given a trained classifier with black box access, 
the attacker tries to infer hyper-parameters of the model using specifically crafted queries~\cite{joon18icrl}. For GPs, possible parameters to be targeted are for example the lengthscale(s) and the chosen covariance function. 

\textbf{Membership inference} describes an attack which aims to learn whether or not some samples were used to train the model~\cite{2016arXiv161005820S,2017arXiv170207464H,DBLP:journals/corr/abs-1806-01246}. Such attacks are generally run in a black box setting, and exploit differences in confidence for trained and unseen data. 
In contrast to deep learning, a GP is not forced to be overly confident on training data, so these attacks are non-trivial.
In our evaluation, we use both confidence (predictive mean) and the predictive variance to deduce this information---a slight variation of known attacks. 

A \textbf{model stealing} attack aims to reproduce the full black-box model~\cite{DBLP:conf/uss/TramerZJRR16,DBLP:journals/corr/PapernotMG16}. For GPs, this amounts to finding out all parameters learned during training and which training data was used, as this information defines the GP completely.
For GPs, this attack is a combination of the previous two attacks.

\subsection{Threat Model}

\begin{table}[!t]
    \normalsize
	\centering
	\caption{Attackers knowledge according to the FAIL model. The symbol $\checkmark$ denotes `known' or `is altered', \text{\sffamily X} the opposite.  }\label{table:attackers}
	\begin{tabular}{@{}l rrrr@{}}
	    \toprule[1.5pt]
		Attacker  & \textsf{F} & \textsf{A} & \textsf{I} & \textsf{L} \\
        \midrule
        Evasion &  $\checkmark$ & \text{\sffamily X} & \text{\sffamily X} & \text{\sffamily X} \\
        Model extraction $l$/lengthscale &  \text{\sffamily X} & $\checkmark$ & \text{\sffamily X} & $\checkmark$ \\
        Model extraction $k$/kernel &  \text{\sffamily X} & $\checkmark$ & \text{\sffamily X} & \text{\sffamily X} \\
        Membership inference & \text{\sffamily X} & $\checkmark$ & $\checkmark$ & $\checkmark$\\
        Model stealing & \text{\sffamily X} & $\checkmark$ & \text{\sffamily X} & \text{\sffamily X}\\
		\bottomrule[1.5pt]
	\end{tabular}
\end{table}

We specify the different adversaries of our empirical study. 
In the FAIL~\cite{suciu2018does} model, \textsf{F} denotes the attacker's knowledge about the features. 
\textsf{A} denotes knowledge about the algorithm applied and \textsf{I} about the training data. \textsf{L} summarizes whether changes to the data by the attacker are constrained. A succinct overview for each attack is given in Table~\ref{table:attackers}.

\textbf{Evasion.} Our attacker knows and changes all features, but is oblivious about the training data and the algorithm. 

\textbf{Model reverse engineering ($l$). } 
The attacker only knows a GPC with an RBF kernel is used. 
The data knowledge varies from black-box to white box, without modifying samples. 

\textbf{Model reverse engineering ($k$). } 
The second attacker only knows GPC is applied. 
Yet, she uses the zero and the ones vector as input, and is thus not constrained on features. 

\textbf{Membership inference. } We assume a worst case scenario, where the attacker obtained a large fraction of data labeled as part of the training set. The attack is not tailored for the learning algorithm, and does not alter the input. 

\textbf{Model stealing. } In our setting, model stealing on a GP can be seen as a combination of the previous two attacks. 


\section{Formal Analysis of Vulnerability}\label{sec:SecGPC}
We take advantage that GP allows a formal analysis. 
First, we show that learning or generalization enables evasion vulnerability on GP.
We then study the interplay of model reverse engineering, membership inference, and model stealing. 

\subsection{Evasion Attacks}
We first define a classifier that cannot be fooled by an adversarial example. 
In the following, we show that a classifier fulfilling this definition, and hence a static security guarantee, is opposed to learning. 
We  briefly define \emph{rejection} of a classifier. 
A classifier can \emph{reject} a sample, in the sense that it does not assign the given sample to any predefined class.

To define a secure classifier, we chose a  
covariance with compact support (for example in \cite{846240}): as the distance from the training data increases, it reaches $0$. Furthermore, there is a $\rho$ such that for all training points, iff point $x'$ is in the closed ball $B(x_i,\rho)$ around a training point $x_i \in X_{tr}$ with radius $\rho$, then $x'$ cannot be an example of another class than $x_i$. In other words, all points in the $\rho$-ball around $x_i$ are of the same class. We formalize the secure classifier 
\begin{equation}
$$ \[ f(x') =  \begin{cases} 
      y_i & \text{iff } x' \in B(x_i,\rho) \\
      \text{reject} & \text{otherwise} 
   \end{cases} \] $$  \label{eq:sec_cla}
\end{equation}
 that cannot be fooled: Changing a sample enough to be classified as a different class means to alter $x'$ so much that $x' \in B(x_j,\rho)$ where $y_i \neq y_j$. 
 Then, by our definition, $x'$ is a valid instance of this other class and not an adversarial example.
%
This secure classifier is equivalent to a GP given the following conditions:
\textit{First}, GP has a rejection option based on $\rho$. \textit{Second},
writing $k(x_i, x_j)$ for the covariance between $x_i$ and $x_j$,
there is no point $x'$ such that for two distinct $x_i$,$x_j \in X_{tr}$ both $k(x_i,x')>0$ and  $k(x_j,x')>0$. 

In other words, we require that GP is able to reject a sample. This can be achieved by setting a threshold on GP's similarity.
Condition two states that the similarity between any two training points is zero, independent of their class. Such a GP, however, has as covariance matrix the identity matrix, as the similarity between any two points is zero. Such a covariance matrix does not allow any learning~\cite{788121}. The details of this equivalence can be found in the Appendix.
Assuming that the second condition does not hold, training points jointly influence classification and the GP generalizes.
\begin{theorem}
Either GP's covariance $K$ is similar to the identity matrix $I$, or $K \neq I$ and learning occurs. Then,  GPR potentially classifies areas outside the $\rho$-balls. Hence, for a test point $x'$ and its corresponding output $p$, $p>\rho$ or $p < -\rho$ although $k(x_i,x')<\rho$, where $x_i$ is the closest training point.
\end{theorem}

\textbf{Proof.} To be classified, we need a classification output $p >\rho$ or $p < -\rho$. We start with the first case, and write 
\begin{equation}
p  \leq \sum_i(\rho - \kappa_i)* [K^{-1}]_i*1 \text{   ,}
\end{equation}
where $[K^{-1}]_i$ is the sum over the inverted covariance matrix column corresponding to training point $x_i$. Before inversion, this column contains the similarities between $i$ and all other training points. So far, we have ignored that we need a test point to obtain this prediction. Without loss of generality, we pick $x'$ which maximizes the sum under the restriction that $x'$ is in none of the $\rho$-balls: hence $\rho - \kappa_i$, and the covariance to $i$th $\rho$-ball is $\kappa_i$. 

There are two cases. In the first, $p \geq \rho$ and we classify outside the $\rho$-ball. In the second, $p=0$ or $0 < p < \rho$. As we choose the maximal $x'$, there are no other points for which $p>\rho$. Then GPR is still secure: no area outside the $\rho$-ball is classified, as the output is below the defined threshold. It remains to be shown, however, that there is no contradiction for the opposite class.  
We proceed analogously with an $x''$ that is chosen to minimize the sum. $\blacksquare$ 

We used in the proof that the minimal output of a point chosen to maximize the sum is zero. Analogously, the maximal value when minimizing the sum is zero as well. This holds due to the abating property of the kernel: As we move away from the data, eventually all similarities become zero, thus the sum is zero as well.
We conclude that generalization 
enables test time attacks such as evasion or adversarial examples. 

%
%
%

\subsection{Attacks against Intellectual Property}\label{sec:fMI}
\begin{figure}[!tt]
\includegraphics[width=\linewidth]{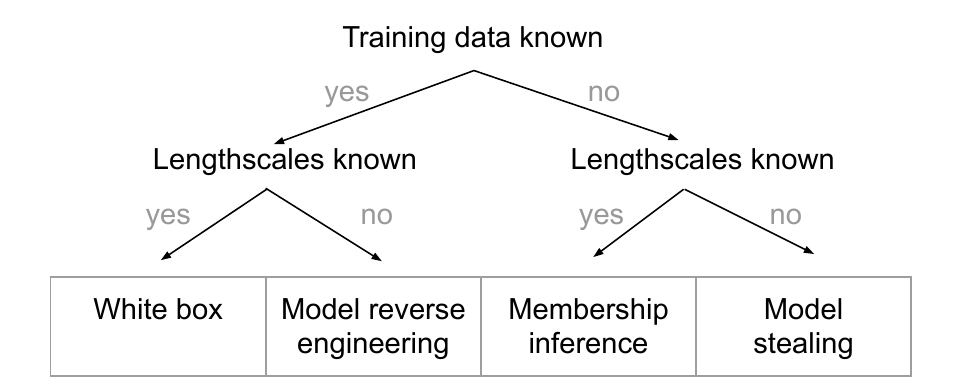}
\caption{The relationship of IP based attacks on GP models.}
\label{fig:stealingComplexity}
\end{figure}

As GPs are an instance of lazy learning, in general all training points and parameters are used during inference. Intuitively, this should ease extraction for the attacker. As we show here, this need not be the case.
We briefly recap the attacker's goal in each attack. In \textbf{model reverse engineering}, she wants to obtain the lengthscale(s), in \textbf{membership inference} the full or partial training data, and in \textbf{model stealing} both lengthscale(s), and full training data. These attacks are strongly related for GPs, as visible in Figure~\ref{fig:stealingComplexity}.

We refresh how classification is computed in GPR (introduced in \cref{pi}). The posterior mean $y^*$ is given as
\begin{equation}\label{eq::outlong}
y^*  = K_{x'}^TK_{\text{tr}}^{-1}Y_{\text{tr}} = \sum_i k(x'-X_i)* K_i^{-1}*Y_i \text{   ,}
\end{equation}
where we iterate over the $n$ training data points. The covariance metric $k$ 
is parametrized using $l$ and $\sigma^2$ when using the RBF kernel.
As lazy learning is used, one might suspect that we can simply \emph{extract} the stored parameters and training data. 
For example, 
independent from the used kernel, we unfold this sum and add the observed output of a GP to obtain an equation system. For simplicity, assume that $Az = y_o $, where $z$ refers to the parameter the attacker wants to retrieve, and $y_o$ is the output observed from the targeted GP. Further $A$ denotes the matrix specified in equation~\ref{eq::outlong}, without $z$. 

The interested reader will have noticed, however, that this equation system solves for unknowns in the number of training points, whereas we need an equation system solving along the features dimensionality. In terms of the above equation, we are actually interested in $A^T z = y_u$, where $y_u$ is an output per feature (where feature and data point dimension are swapped, or $X^T$).
Hence, $y_u$ is not an output for any GP trained on $X$: it corresponds to a label per feature.
In the original task, the features are ``lost'' in the kernel Hilbert space inside the GP, and the attacker has no equation system since there is no $y_u$.

The existing equation system, can only be used to determine the lengthscale iff there is only one global lengthscale set, and the GP has no other unknown parameter. Otherwise, the equation system is not properly specified, and no analytic computation is possible. 
We thus conclude that lazy-learning, albeit counter-intuitively, is not less privacy resilient that other classifiers. 
Instead, however, the attacker can take advantage that GP are deterministic. A GP with the same parameters and data always yields the same output. In the following, empirical section we evaluate this type of attack.

\subsection{Conclusion}
GP, as it learns, is vulnerable to evasion attacks. Concerning IP-related attacks, we can exclude the possibility that the attacker analytically determines training data or lengthscales, with the exception of a single learned parameter for all dimensions (for example in a linear kernel).


\section{Empirical Study of Vulnerability}
We now describe our complementary empirical study.
We start with the setting including data-sets, implementation, and parameters. Afterwards, we detail the results on evasion, model reverse engineering and membership inference.

\subsection{Experimental Setting}
We first describe the general setting. Specifics are 
 given jointly with the corresponding attacks. 

\textbf{Data and implementation.} Our study encompasses several data-sets, including security tasks such as Malware (Hidost, Drebin)~\cite{hidost,arp2014drebin} and Spam detection~\cite{Lichman:2013}. Additionally, we investigate fake banknote detection by~\cite{Lichman:2013}, the MNIST benchmark data set~\cite{lecun-98}, and the SVHN data set~\cite{37648}. 
We use Python and GPy for the Gaussian Process approaches~\cite{lawrence2004gaussian}. We show further information on the trained GPCs in Fig. \ref{table:data-settings}, such as the number of training samples and lengthscales used and achieved accuracies.
To obtain adversarial examples, we use Tensorflow~\cite{tensorflow2015-whitepaper} and the Cleverhans library 1.0.0~\cite{DBLP:journals/corr/GoodfellowPM16} for DNN, and other public implementations~\cite{DBLP:journals/corr/CarliniW16a,2017arXiv171106598G}.

\textbf{Parameter choices.} We train our GPC using the RBF kernel with a predefined lengthscale. This GPC is fitted until convergence or for $100$ iterations. 
For each task, we chose two lengthscales that achieve similar accuracy (see Table~\ref{table:data-settings}). More details on how we determined the two used lengthscales can be found in the Appendix. 


\begin{table}[!t]
	\centering
	\small
	\caption{Number of samples used in training $n$, lengthscales $l$ and accuracies (rejection  if $\mu = 0$, written Acc$_r$). }\label{table:data-settings}
	\begin{tabular}{@{}l rrrrrrr@{}}
	    \toprule[1.5pt]
	    	& & \multicolumn{3}{c}{short $l$} & \multicolumn{3}{c}{long $l$}  \\
	    	\cmidrule(l){3-5} \cmidrule(l){6-8} 
	    Data-set & $n$ & $l$  & Acc$_r$& Acc & $l$ & Acc$_r$ & Acc \\
        \midrule
        Hidost & $500$ & $.5$ &$98.4$&$98.4$& $1.9$ & $97.7$ & $99.6$  \\
        Drebin & $750$ & $.5$ & $54.4$ & $94$ & $1.9$ & $94.8$ & $94.8$  \\
        Spam & $500$ & $.3$ & $92.6$& $91.7$ & $5$ &$92.7$ & $90.2$  \\
        Bank & $500$& $.3$ & $100$&$100$& $2$ & $100$ & $100$ \\
        MNIST91 & $500$ & $1$ & $98.9$&$98.3$& $8$ & $99.5$& $99.5$  \\
        MNIST38 & $500$ & $1$ & $93.4$&$93.4$& $8$ & $97.4$ & $97.1$ \\
        SVHN91 & $1500$ & $8$ & $85.4$& $88.5$& $16$ & $83.8$ & $87.6$ \\
        SVHN10 & $1500$ & $8$ & $88.7$& $88.7$ & $16$ & $88.7$ &$88.7$ \\
		\bottomrule[1.5pt]
	\end{tabular}
\end{table}

\subsection{Evasion / Adversarial Examples}\label{sec:lenvuln}
We expect that a GP with a long lengthscale misclassifies fewer adversarial examples: A larger perturbation $\delta$ is needed to cause the same change in the output. 

\textbf{Setting.} 
To obtain adversarial examples independent of the specific curvature, 
we do not craft on the GPCs tested.
We instead transfer FGSM, JSMA and $\mathbf{L_x}$ attacks from deep neural networks, linear SVM and a GPC substitute. 
Our intention is to study a wide range of attacks, including optimized, unoptimized, one-step and iterative attacks as well as different metrics ($L_0$, $L_2$, and $L_\infty $). We summarize all attacks based on the Jacobian in JBM, sort FGSM according to $\epsilon$ and plot the $\mathbf{L_x}$ attacks according to the norm optimized (for example $L_2$ for the $L_2$-norm attack). 

We compare how well the previously chosen lengthscales recover the correct class when facing adversarial examples. 
In our plots, a value above zero denotes that the shorter lengthscale classified more data correctly, where the numbers are difference in absolute percent. Below zero, a longer lengthscale (flat curvature) performed better.

 \begin{figure}[t]
\begin{subfigure}[b]{1.0\linewidth}
\includegraphics[width=\linewidth , trim={1cm 0.5cm 1cm 0},clip]{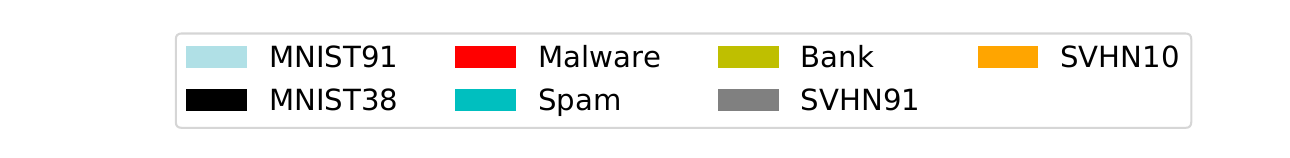}
\end{subfigure}
\centering
\begin{subfigure}[b]{0.85\linewidth}
\includegraphics[width=\linewidth]{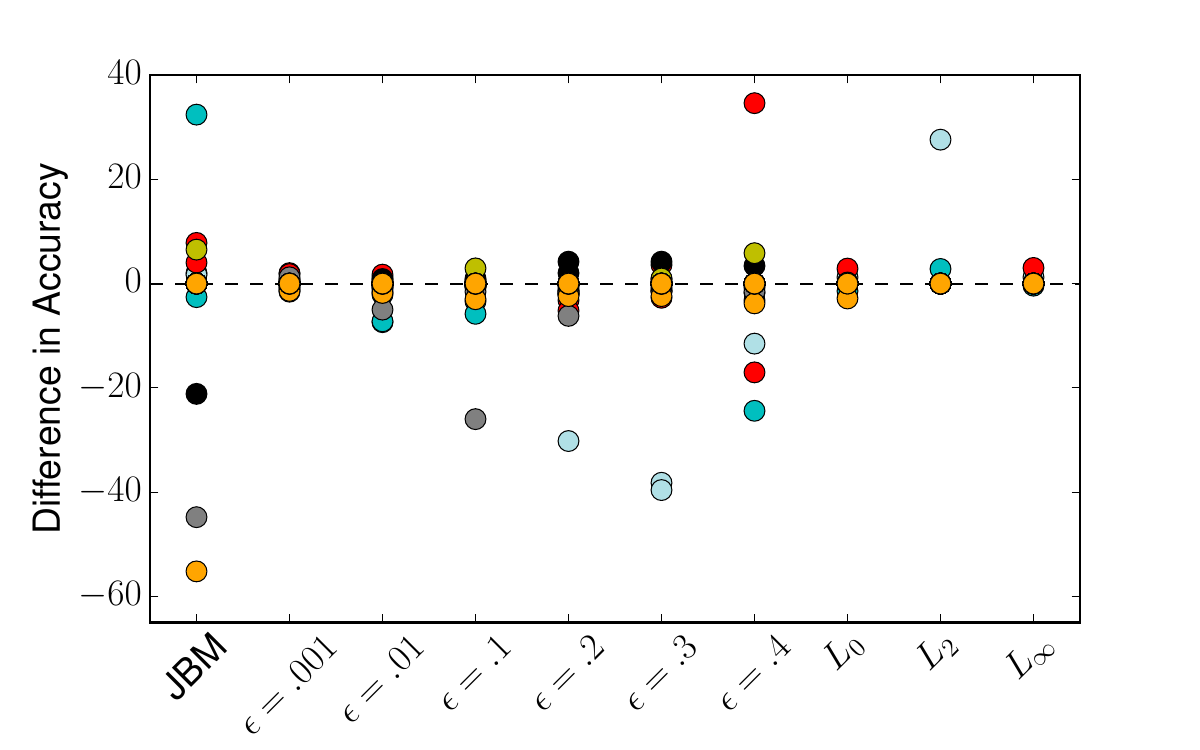}
\end{subfigure}
\caption{Vulnerability and Curvature in GPC. Above zero denotes that more examples are correctly classified by a GPC with long $l$, below zero with short $l$.}\label{fig:lengtscVuln}
\end{figure}

\textbf{Results.} We plot the results of our experiments in Fig. \ref{fig:lengtscVuln}. A short lengthscale generally classifies more adversarial examples as their original class. In particular on $L_\infty$ attacks (with $\epsilon > 0.01$), a short lengthscale performs better. A long lengthscale is advantageous for optimized attacks like  $L_2$. 
%
%

\begin{figure}
\begin{subfigure}[b]{\linewidth}
\includegraphics[width=0.98\linewidth]{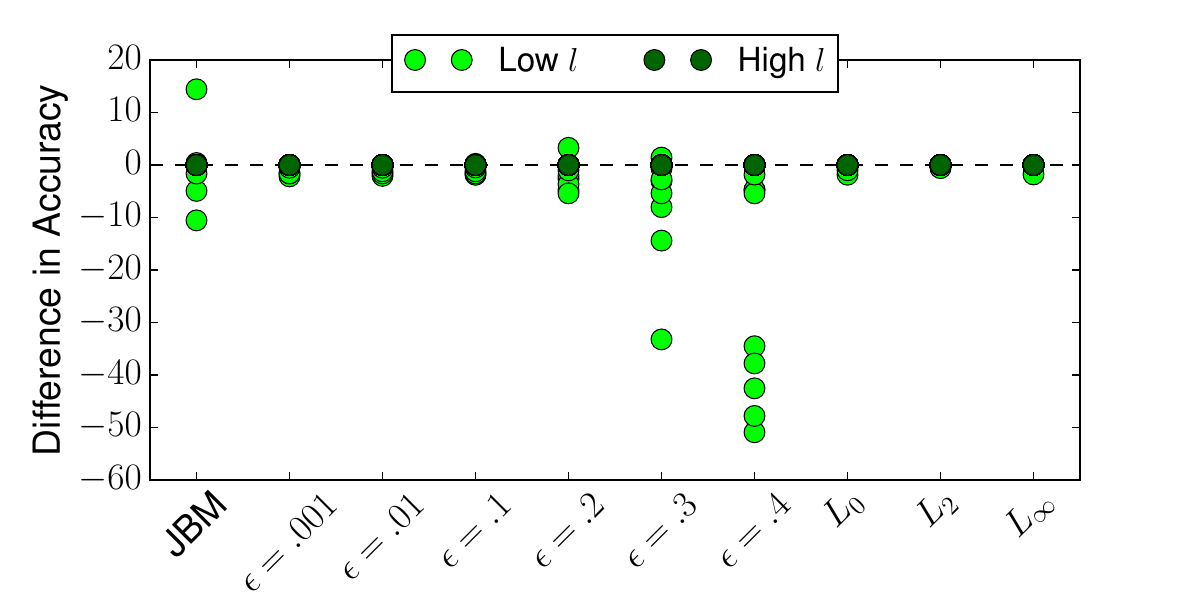}
\end{subfigure}
\caption{Vulnerability, lengthscale and rejection option in GPC. Above zero denotes that more examples are correctly classified or rejected by a GPC without a rejection option.}\label{fig:rejectLengthscale}
\end{figure}

We also investigate how lengthscale affects rejection, as our preliminary results show only a slight advantage for steep curvature GPs without rejection.
In Fig. \ref{fig:rejectLengthscale}, a negative number denotes how much absolute percent the reject performs better compared to a classifier without reject.
A positive number means that accuracy for rejection is worse.
There is no difference in vulnerability to evasion for a long lengthscale.
 For a short lengthscale, the effect is positive or neutral, with only two negative cases. These two cases stem from the highly imbalanced Hidost data set. By chance, the assignment of the forced classification was in favor of the larger class.     

\textbf{Conclusion.} 
Only classifiers with steep decision functions benefit from rejection. 
We hypothesize that a short lengthscale allows for larger areas where the rejection area is actually used, whereas a long lengthscale leads to confident classification in areas where no benign data was seen.

\subsection{Model Reverse Engineering}
Model reverse engineering refers to the retrieval of hyper-parameters of the model. We introduce two new attacks to reverse engineer GP's lengthscale and kernel.
 
\textbf{Setting (lengthscale).} 
We pick the same lengthscales as before and evaluate whether the attacker is able to determine the lengthscale of a target GP. The attack is a binary search to obtain $l$. The distance between the outputs of two GPs shrink as the lengthscale chosen by the attacker, $l_a$, approaches the original lengthscale $l$. We evaluate three settings: Training GPC on the same data as the victim, mixed (half/half) and disjoint data. In each setting, we train $50$ GPCs, starting with a lengthscale $l_{a} = l/2$ and increasing the lengthscale in steps of $(1/50)l$. We then compute the absolute difference between the outputs of the  GPCs on hold out, unused test data. 

\begin{figure}[t]
\begin{subfigure}[b]{\linewidth}
\centering
\includegraphics[width=\linewidth , trim={2cm 0.5cm 2cm 0},clip]{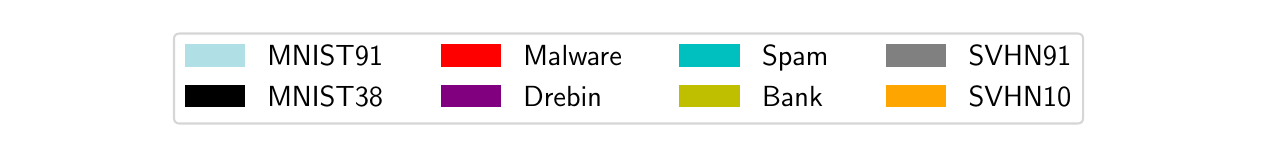}
\end{subfigure}
\newline
\begin{subfigure}[b]{0.495\linewidth}
\includegraphics[width=\linewidth]{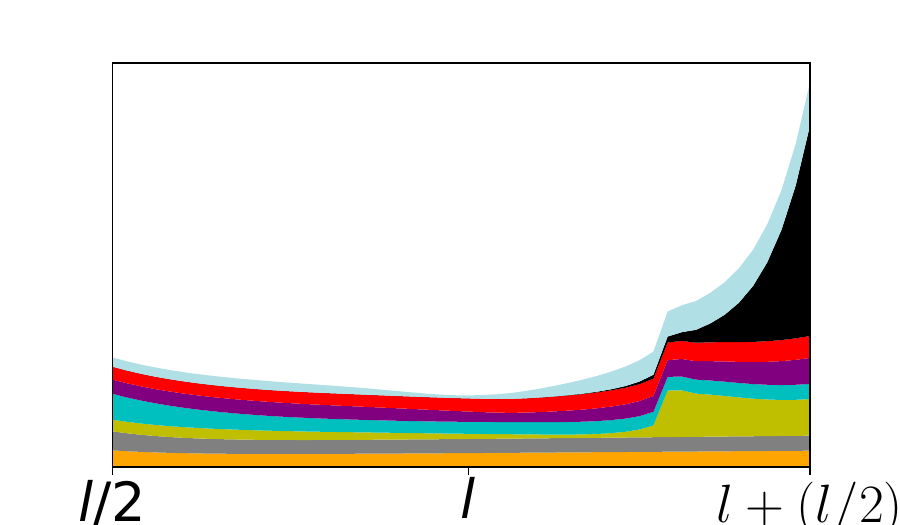}
\caption{Short $l$, mixed data.}
\end{subfigure}
\begin{subfigure}[b]{0.495\linewidth}
\includegraphics[width=\linewidth]{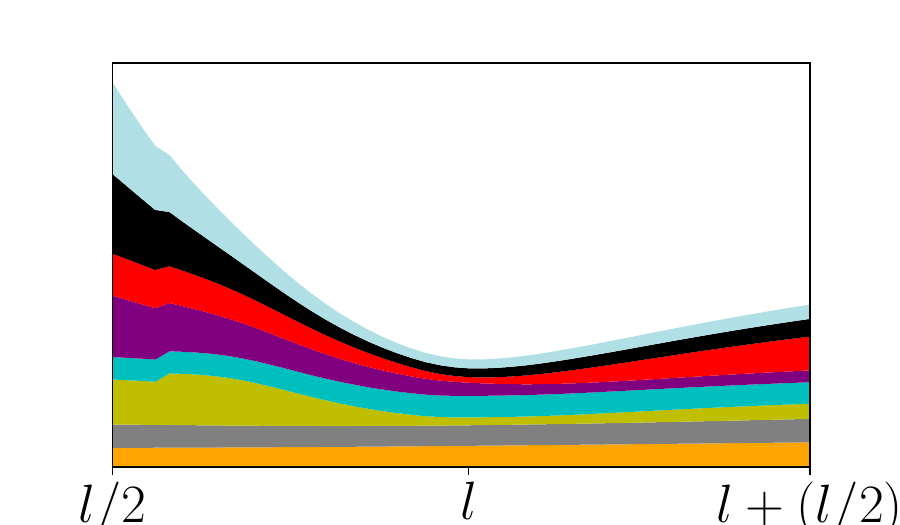}
\caption{Long $l$, mixed data.}
\end{subfigure}
\begin{subfigure}[b]{0.495\linewidth}
\includegraphics[width=\linewidth]{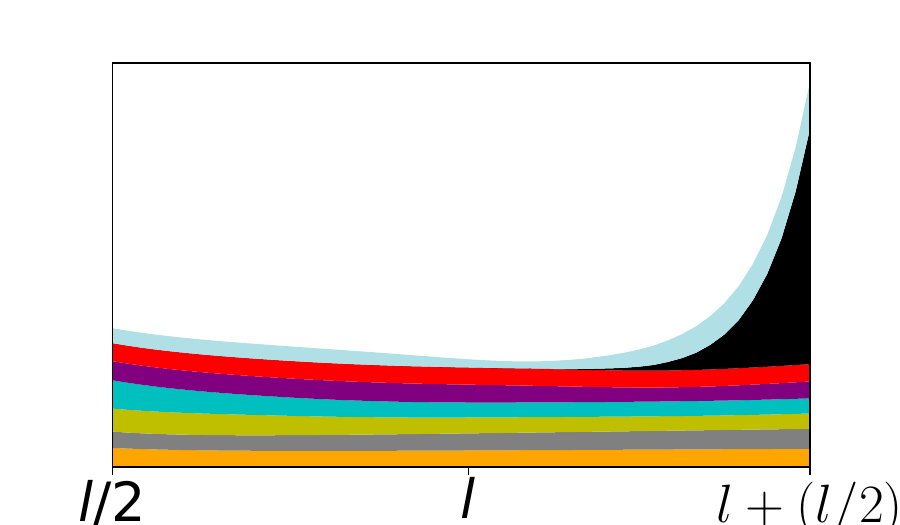}
\caption{Short $l$, disjoint data.}
\end{subfigure}
\begin{subfigure}[b]{0.495\linewidth}
\includegraphics[width=\linewidth]{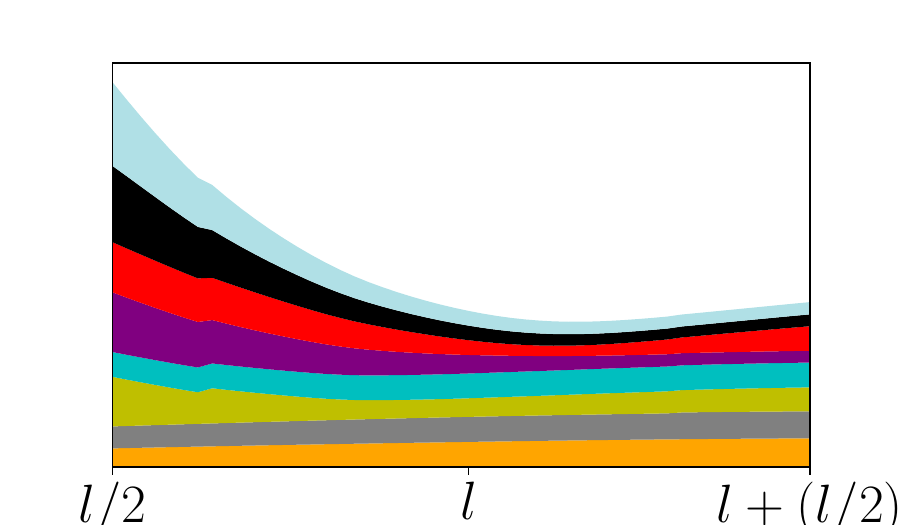}
\caption{Long $l$, disjoint data.}
\end{subfigure}
\caption{Normalized, absolute differences in output for different data sets when binary searching a GP's lengthscale. $x$-axis is $L_a$; hence at $l$, lengthscales are equivalent.}\label{fig:ModelStealing}
\end{figure}

\textbf{Results (lengthscale).} 
As GPs are deterministic, all distances decrease towards the original $l$ when the training data is fully known. We thus omit plotting these results,  and study the more interesting cases of mixed data.

For mixed data (upper plots of Fig. \ref{fig:ModelStealing}), the results are less clear. In general, distances decrease towards $l$. Given a long lengthscale, the distances are smallest around $l$. An exception are SVHN and Spam, where the distances remain constant for all $l_a$s. On Drebin, the distances are smallest around $l+(l/2)$. The results vary for a short lengthscale: for some data sets (MNIST91, Bank) the distance is closest to $l$, For others (including SVHN and Malware), the smallest distance is $l/2$.   

In case of disjoint data sets (bottom plots of Fig. \ref{fig:ModelStealing}), the results are even less pronounced. The distance slightly decrease towards the original lengthscale, yet the average minimum is at a lengthscale $>l$. In case of a short lengthscale, there are no differences at all. An exception are the two MNIST tasks, where however again the minimum is $>l$.  

In general, a lengthscale can be approximated using binary search. More concretely, the estimate is close when the original lengthscale is long: The difference to the original lengthscale is then between $0.006l$ and $0.008l$. This corresponds to wrongly estimating  the largest lengthscale of SVHN by $1.28$ ($17.28$ instead of $16.0$) or the smallest (Bank) by $0.16$ (estimating $2.16$ instead of $2.0$). For a short lengthscale, the estimate for MNIST91's lengthscale is around $1.04$ instead of $1$. For cases except MNIST91, the estimate is inaccurate or indeterminable. 

\begin{figure}[!t]
\centering
\includegraphics[width=\linewidth]{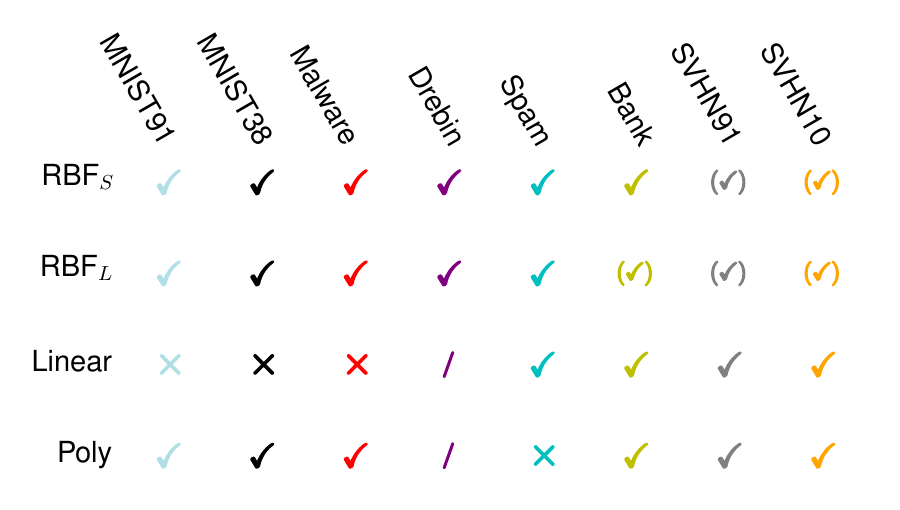}
\caption{Stealing the kernel of a GPC (columns), \checkmark denotes successful extraction. $\times$ denotes a failed attack, (\checkmark) that the attack succeeded only in an easy-to-defeat variant. Some cases were not evaluated ($/$) as test accuracy was too low.}\label{fig:kernel}
\end{figure}

\textbf{Setting (kernel).} 
The goal of the attacker is to determine the kernel used in a black-box GPC. 
We assume the victim uses one of the following kernels, RBF (with the same lengthscales as before), linear, or polynomial.
We exclude the results on Drebin with the linear and polynomial kernel, as their accuracies are close to a random guess.

\textbf{Attack Description (kernel).} 
An RBF-kernel will output close to zero far away from the seen data. 
Hence, we input the target GPC a zero and ones sample and deduce an RBF-kernel is used if the output is close to $0.5$.
We also use a more extreme, easy to defeat variant of this attack where the given samples contain only features equivalent to $\pm 10$.
To preserve feature meaning, we could also compute an \emph{unusual}, far away sample. Due to the diversity of our data-sets, we leave this variant for future work. 

In case neither output is $0.5$, we run a second round of queries.
We assume a linear kernel is limited in its expressivity, and leads to less confidently classified data. We thus submit a batch of test points, and classify a kernel as polynomial (nonlinear) if the distribution of outputs is bi-modal with most values scattered around 0 and 1.
We hence compute the mean of the values above and below $0.5$.
The threshold for the decision is that both means are further apart than $0.7$.
This threshold was determined on the additional credit data-set~\cite{Lichman:2013}, which is otherwise not used in this evaluation.

We train different GPCs, each using a different kernel (RBF kernel with several learned lengthscales, linear, polynomial kernel). The attacker determines, with the above heuristics, the used kernel. Our results are depicted in Fig.~\ref{fig:kernel}.

\textbf{Results (kernel).} In the majority of cases, the attack succeeds independent of lengthscale or kernel used. In three of eight cases, the linear kernel is wrongly determined as nonlinear, indicating that it confidently classified the data against our expectation. In one case, on the spam data, the polynomial kernel is wrongly determined as RBF kernel. There are also some cases on the Bank and SVHN data-sets where the RBF kernel is only correctly predicted if we use the $\pm 10$ filled samples. Otherwise, the attacker's classification is that the victim uses the polynomial kernel on bank, or the linear kernels on the SVHN tasks.

There are very few differences between using full test data, 500, 50, or as few as 10 samples for the linear/nonlinear query. Only on the Bank data-set we observed that the linear kernel was classified as polynomial kernel using 50 samples or less. All other results remained consistent.


\textbf{Conclusion.} 
Empirically, the lengthscale can be recovered easily if the training data is (partially) known. This relates to the GP being deterministic. 
Otherwise, the attacker can reasonably well approximate the lengthscale given that the targeted GPC has a long lengthscale.
We hypothesize that the long lengthscale is easier to extract as it is less prone to small changes in the data distribution.
The kernel is, instead, easy to deduce independent of the lengthscale. 
Our attack currently fails if the linear kernel fits the data well (MNIST, Malware) or the polynomial kernel's decision boundary passes the origin or the ones vector. 
With a long lengthscale, the RBF kernel (Bank, SVHN) outputs relatively large values even far away from the data.
 Yet, we find no absolute value for this to happen.
Another natural defense to our attack are custom-based kernels. 
We leave this cases for future work, and conclude that our heuristic works well for the given data-sets and the kernel set $\{$RBF, linear, polynomial$\}$.

\subsection{Membership Inference}
We investigate how good an attacker can determine which points were used in training. First, we study the general setting. Afterwards, we investigate particular settings influencing the attackers success: overfitting, distribution drift, and sparsity. 

To study a worst case scenario, the attacker has an oracle that labels a large fraction of the training data as such. This attacker is slightly stronger than the shadow models used in~\cite{2016arXiv161005820S}. The attacker trains a fresh classifier that predicts membership for unseen data points.

\textbf{Setting.}  
The target GPs are trained using the same lengthscales as before. We then build a data-set using the output of the GPs and membership labels indicating if a data point was used in training. The data-set is split randomly in test data ($50$ points) and training data (the remainder). The training data is used to train a fresh classifier. We tested DNN, decision trees, random forests and AdaBoost classifiers. As the random forest classifier performed consistently best, we applied only random forests. We report accuracy and random guess on the test data. 

\begin{figure}[t]
\begin{subfigure}{\linewidth}
\includegraphics[width=\linewidth]{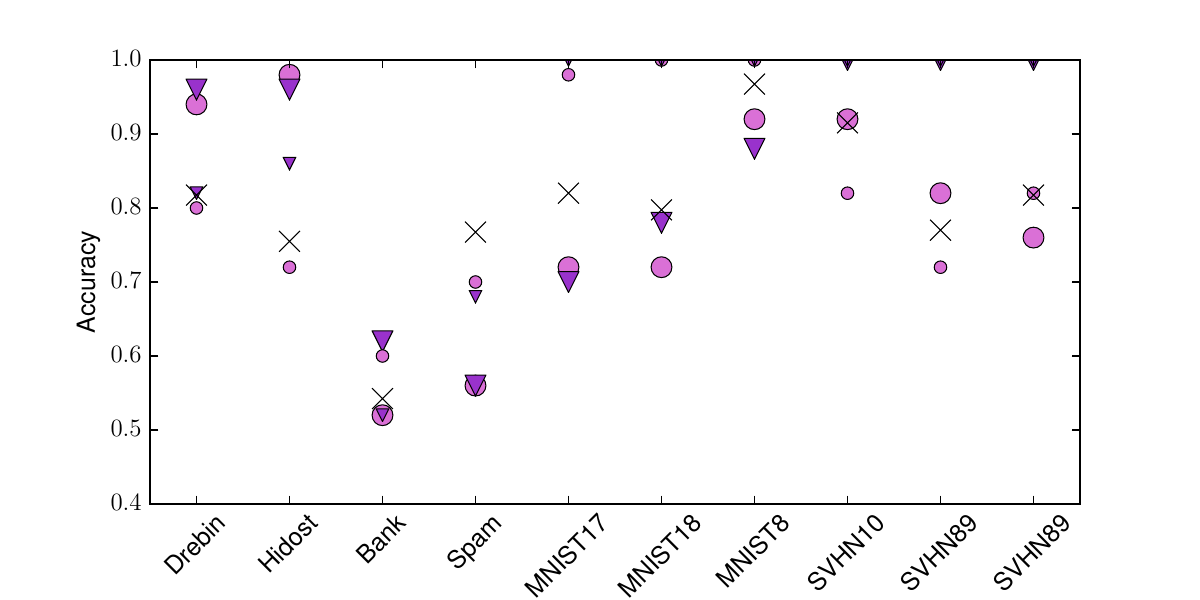}
\caption{Classifier trained on variance (triangles) or mean (dots).}\label{fig:MIa}
\end{subfigure}
\begin{subfigure}{\linewidth}
\includegraphics[width=\linewidth]{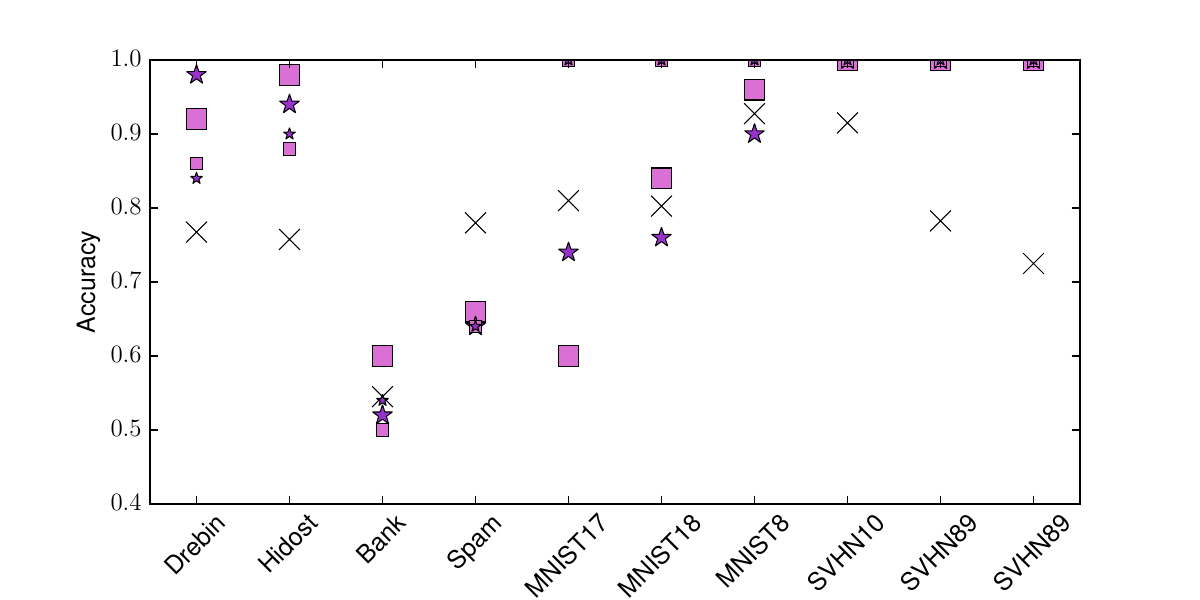}
\caption{Training on mean and variance (square) or latent mean (star).}\label{fig:MIc}
\end{subfigure}
\caption{Lengthscale and membership inference on GPC. Bigger symbols denote a long, small symbols a short lengthscale of targeted GPC. x denotes random guess.}\label{fig:lengtscaleMI}
\end{figure}

\textbf{Results.} 
We train the random forests on predictive mean (dots), variance (triangle) (both Fig. \ref{fig:MIa}), mean and variance (squares), or the unnormalized, latent mean (stars) (Fig. \ref{fig:MIc}).
Overall, using only the predictive mean and a long lengthscale (larger markers), no data set is vulnerable, with the exception of the two Malware data sets.
For mean and variance and latent mean settings, 
the attacker succeeds in both cases on all SVHN tasks or when using a small lengthscale, with the exception of non-vision tasks. The attacker is also  successful on the Malware data sets with a long lengthscale.

On the bank and spam data, the attack is never successful. In general, a shorter lengthscale is more vulnerable. On the Malware data sets, the inverse holds: here, a short lengthscale benefits the defender. Before we focus on these cases, however, we investigate what enables the attacks on the SVHN data and why a short lengthscale is beneficial for the adversary.

\textbf{Overfitting, distribution drift, and sparsity.} 
We compare training and test accuracies to measure \textbf{overfitting}. 
On the bank data, training and test accuracy are both 100\%.
On all other data-sets, the difference between test and train accuracy is smaller for a long lengthscale.  
Hence, slight overfitting occurs at short lengthscales, and enables  membership inference.

To analyze \textbf{distribution drift}, we measure the standard deviation over the distances between training and test data. 
As GP adapts the similarity during training,
we expect the test data to cause larger variance in the distance if the data is distributed differently. 
All SVHN and MNIST8 settings with a small lengthscale show a variance two magnitudes larger between test and training data than among either.
Thus, the attack was enabled as training and test data were different from the perspective of GPC. This might imply that the model is not expressive enough to model the data in detail.

Two cases of successful membership inference are left unexplained: the Malware data-sets, Hidost and Drebin. We suspect that \textbf{sparsity} causes the vulnerability. The average percentage of features $>0$ on the full data-set is $<0.001$\% $\pm0.0006$ on Drebin and $\sim 12$\% $\pm3.8$ on Hidost. Next is MNIST( 1 vs 7 with around $14$\% $\pm 4.1$, 1 vs 8 with $\sim 16$\% $\pm6.5$ and 8 with $\sim18$\% $\pm5.2$). All other data-sets exhibit less sparsity ($>20$\%, Spam) or well above $70$\% (all remaining data-sets). 
 
 The difference in sparsity between Hidost and MNIST is small, yet the discrepancy to robust data-sets (Bank, Spam) is large. 
 To account for sparsity, we apply a GPC using inducing variables (GPy's sparse GPC). 
Such a GPC also optimizes over the training points: the training data is then not directly stored.
 
\begin{figure}[t]
\centering
\begin{subfigure}[b]{0.3\linewidth}
\includegraphics[width=\linewidth]{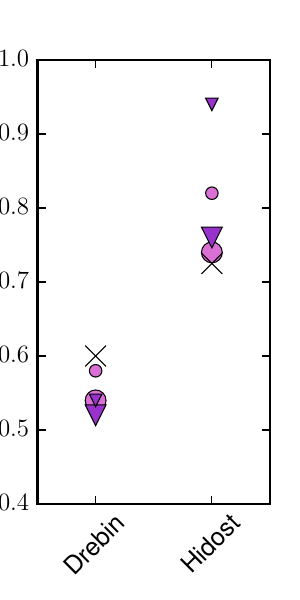}
\end{subfigure}
\begin{subfigure}[b]{0.3\linewidth}
\includegraphics[width=\linewidth]{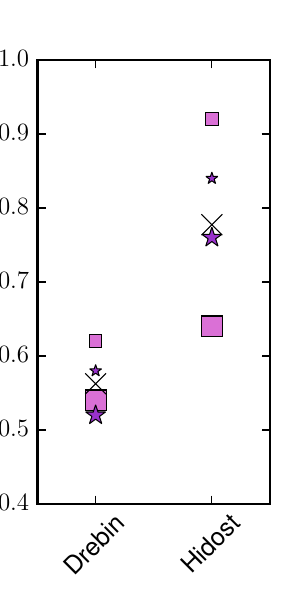}
\end{subfigure}
\caption{Accuracy of membership inference on a sparse GPC. Bigger symbols denote a long, small symbols a short lengthscale on targeted GPC. x denotes random guess. Left plot: Classifier trained on variance (triangles) or mean (dots). Right: Training on mean and variance (square) or latent mean (star).}\label{fig:lengtscaleMISparse}
\end{figure}

In Fig.~\ref{fig:lengtscaleMISparse}, we investigate the same settings from the previous study. The attacker's accuracy is now on all settings close to a random guess, with the exception of a short lengthscale for Hidost on mean or variance, latent mean, or mean and variance. For Drebin, we observe a very small improvement over random guess when a short lengthscale is used and the attacker accesses the GP's predictive mean and variance.  

\textbf{Conclusion.} 
Even under a strong attacker, membership inference is not successful when there is no distribution drift, overfitting is properly taken care of, 
or a sparse GP with a long lengthscale is applied.
Robustness of GPs towards membership inference is somewhat expected,
as a GP is not required to be overly confident on training data.
The effect of the lengthscale is also intuitive. A short lengthscale allows each training point only local influence, easing inference about membership.
With a long lengthscale, each point influences a large area, making it harder to locate the exact training point.


\section{Conclusion}
We investigated the security of GPs at test time towards evasion, model reverse engineering, membership inference, and model stealing. We conclude that attack vectors on classification should not be seen in isolation, as a mitigation towards one attack might enable or ease another attack. 

Formally, we show that evasion is enabled by learning, and any learned GP is vulnerable. 
Against possible intuition, lazy learning is not per se more vulnerable towards IP attacks. Still, a re-computation of the lengthscale is possible
if kernel and the training data are fully known.
Yet, no further parameters can be analytically retrieved from given output.

We also study empirical vulnerability, and leveraged the property of a GP to fit a model with a predefined decision 
curvature.
Our study encompasses six data-sets. 
Summarizing, a short lengthscale leaks the data, and is vulnerable to optimized evasion attacks. 
A long lengthscale leaks the parameters of the GP, and is vulnerable to one-step attacks with large $\epsilon$.
The kernel can be determined independent of the used lengthscale.
We conclude that attacks on classification should not be studied in isolation, but in relation to each other.

\section*{Acknowledgments}
The authors would like to thank the anonymous reviewers, Christian Rossow, Thomas A. Trost, and David Pfaff for their helpful feedback. 
This work was supported by the German Federal Ministry of Education and
Research (BMBF) through funding for the Center for IT-Security,
Privacy and Accountability (CISPA) (FKZ: 16KIS0753). 
This work has further been supported by the Engineering and Physical Research Council (EPSRC) Research Project EP/N014162/1. 

\bibliographystyle{IEEEtran}
\bibliography{IEEEabrv,advgp} 

\newpage
\appendix


\textbf{GP as a secure classifier.}
We show that given the previous two conditions from the main paper, GP is indeed equivalent to a secure classifier.
To start, we recap the two assumptions.
\textit{First}, GP has a rejection option based on $\rho$, and \textit{second},
there is no point $x'$ such that for two distinct $x_i$,$x_j \in X_{tr}$ both $k(x_i,x')>0$ and  $k(x_j,x')>0$. 
In other words, the GP can reject a sample, and we assume that the used covariance is $0$ between any two training points.
the predictive mean of GPR is composed of weighted labels, where the weight is the similarity between the training points and the queried test point, as formalized in \cref{pi}.
Throughout this analysis, we assume the values representing the two classes to be $1$ for class $1$ and $-1$ for class $0$ respectively. 

When assumption two holds, the kernel matrix is the identity matrix: the similarity between any two points is zero, and the similarity of each point with itself is $1.0$. A test point $x'$ is at most close to a single training point, here $x_{n}$. We summarize:

\begin{equation}
y^*=k(x',x_{n})*1*y_{n} \text{   ,}
\end{equation} 
where $1$ denotes the similarity from $x_{n}$ to itself; all other terms are zero as $k(x',x_{i})=0$ iff $i \neq n$. This classifier is almost equivalent to \cref{eq:sec_cla}, as it classifies each point if it is in the $\rho$-ball of any training point. 

To fulfill the equivalence, we define rejection for GPR (assumption 1). We need to reject any test point for which $k(x',x_{n})\leq \rho$. The minimum and maximum output of GPR is (due to the labels) either $1 = y_1$ or $-1 = y_0$, thus we do not classify a sample iff
$y^* \in [-1+\tau_0,1-\tau_1]$,
where we chose $\tau_0 = \tau_1 = \rho$ to reject points further away than $\rho$. We may, however, define a different thresholds (where $0.0 < \tau < 1.0$) for the two classes or even each test point, if required. 

\vspace{3em}

\begin{figure*}[bp]
\begin{subfigure}[b]{\linewidth}
\includegraphics[width=0.9\linewidth]{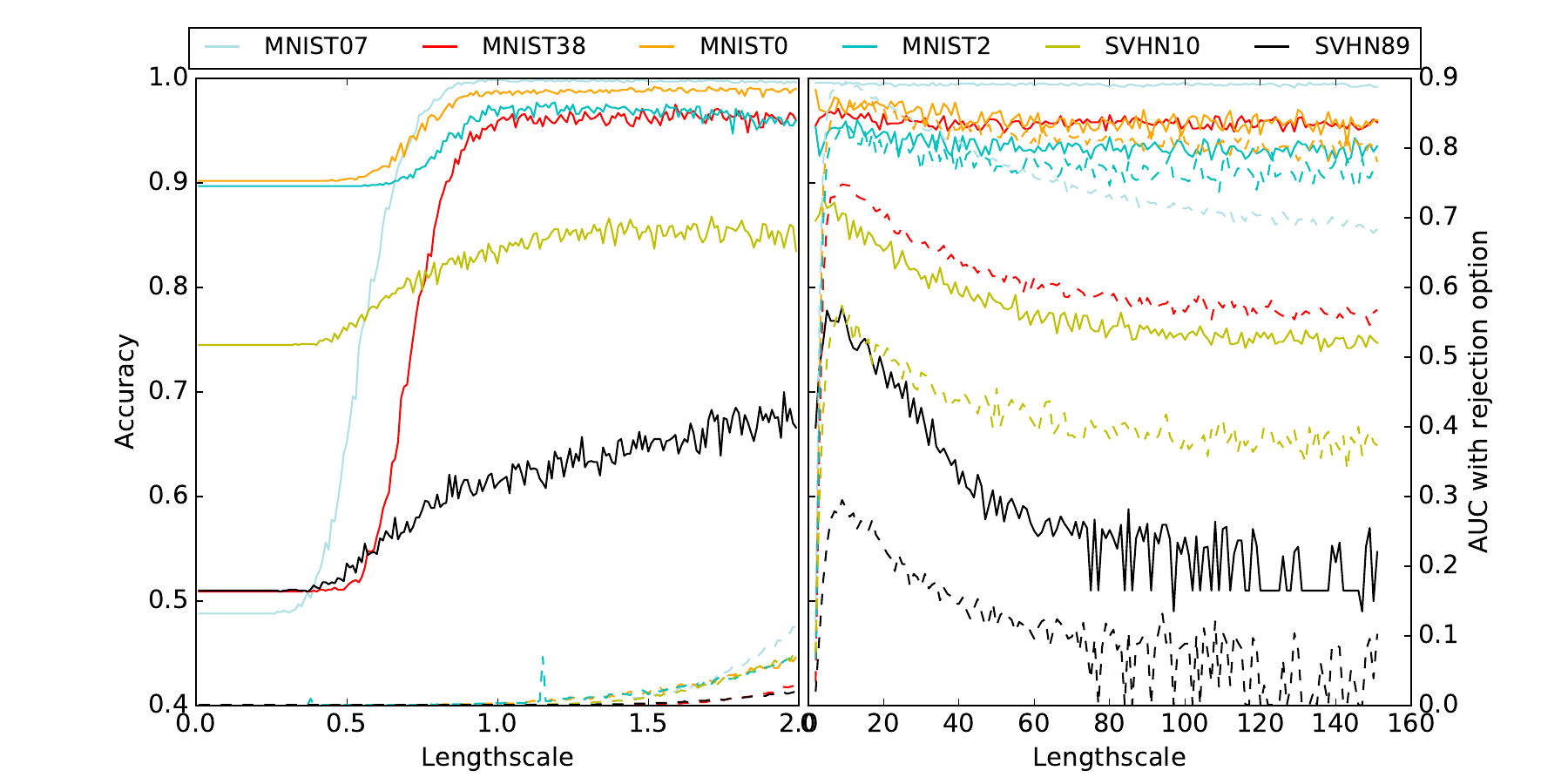}
\end{subfigure}
\begin{subfigure}[b]{\linewidth}
\includegraphics[width=0.9\linewidth]{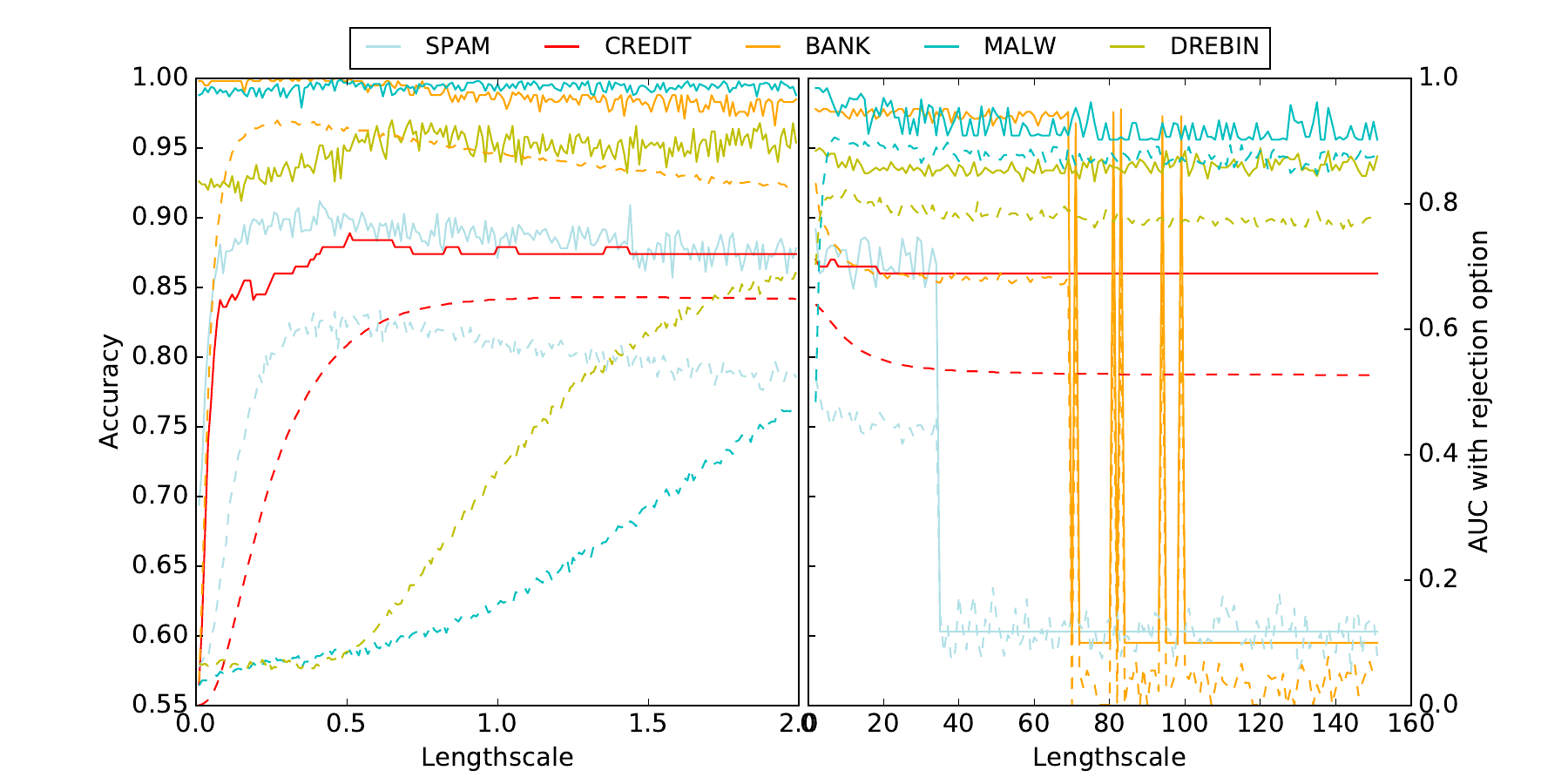}
\end{subfigure}
\caption{Accuracy and variations of lengthscales in GP. Accuracies are lines, AUC when using rejection is depicted as dotted lines. }\label{fig:lengtscaleAll}
\end{figure*}

\textbf{Lengthscale Experiments.}
To determine the two lengthscales used throughout the paper, we trained a range of GPCs on each data set. We varied the lengthscale in small steps between $0.0$ and $2.0$ in steps of $0.01$ and between $2.0$ and $151.0$ in steps of $1.0$. We report the resulting accuracies and AUC with all rejections thresholds in \cref{fig:lengtscaleAll}. The figure include an additional data set Credit, which was later excluded as we found the accuracy to varied strongly (by $10$\%) if the split between test and training data was changed.

\end{document}